\documentstyle[12pt,titlepage]{article} 
\title{Long Distance Contributions to B-Decays into Higher K-Resonances}
\author{Mohammad R. Ahmady and Dongsheng Liu}
\date{January, 1993}   
\def\_{\rule{.3em}{.15ex}}  

\begin{document}
\begin{titlepage}
 \begin{center}
  \vspace{0.75in}
  {\bf {\LARGE Non-Leptonic B Decays into K-Resonances} \\
  \vspace{0.75in}
  Mohammad R. Ahmady} \\
  Department of Applied Mathematics, The University of Western
Ontario \\
  London, Ontario, Canada \\  {\bf Dongsheng Liu}\\
  Department of Physics, University of Tasmania \\
  Australia  \\
  \vspace{1in}
  ABSTRACT \\
  \vspace{0.5in}
  \end{center}
  \begin{quotation}
  \noindent We estimate the non-leptonic B decays $B \rightarrow
(\psi ,\psi
  ^\prime , \chi_{1c})+K^i$, where $K^i$ are various K-meson
resonances.  We
  use the model of Isgur, Wise, Scora and Grinstein in the context of
heavy
  quark effective theory, to calculate the hadronic matrix elements.
Our
  estimates show that a substantial fraction of $B \rightarrow X_s
\psi$
  results in higher resonances of K-meson and besides $B \rightarrow
K(K^*)
  \psi$, a considerable fraction of $B \rightarrow X_s (c\bar c)$
goes to
  $B \rightarrow (K,K^*)+( \psi^{\prime},\chi_{1c})$.

  \end{quotation}
  \vspace{0.5in}

\end{titlepage}

processed

\newcommand{\da}{\mbox{$\scriptscriptstyle \dag$}}
\newcommand{\lag}{\mbox{$\cal L$}}
\newcommand{\tr}{\mbox{\rm Tr\space}}
\newcommand{\fc}{\mbox{${\widetilde F}_\pi ^2$}}
\newcommand{\ns}{\textstyle}
\newcommand{\si}{\scriptstyle}

\section{Introduction}

Recently, there has been some progress in the measurement of some B
decays
to final states containing strange quarks \cite{CLEO,HFC}.  The
accurate measurement
 of rare B decays can shed some new light on the Standard Model as
well as
 the physics beyond it.  On the other hand, nonleptonic B decays like
$B
 \rightarrow K_s J/\psi$ could prove to be important for CP violation
\cite{NQ}.  It
 is the latter processes that are the main focus of this paper.

 The CKM favored nonleptonic B decays have been extensively
investigated in
 the literature.  In fact, the two body decay $B \rightarrow K^{(*)}
J/\psi$
  is used for the measurement of the B life time \cite{OPAL}.  On the
other hand,
   the same process could generate a significant background to the
rare decay
    $B \rightarrow K^* \gamma$ through $J/\psi -\gamma$ conversion
\cite{ATL}.
  Also, a match between theoretical estimate and
experimental
  value of this decay could be used as a check for spectator model.

The main source of uncertainty in the theoretical calculations, is
the
evaluation of the hadronic matrix element.  In a previous paper on
this
subject, we used the new symmetries in the heavy quark limit to
calculate
these matrix elements in terms of a single universal Isgur-Wise
function (to
the leading order in heavy quark expansion) \cite{AL}.  In that
paper, we
concluded that indeed this method could be used for an estimate of
these
processes even though the {\bf s} quark is not particularly heavy.

On the other hand, due to the large mass difference between B meson
and $J/\psi
, \psi^\prime {\rm and} \chi_{1c}$ there is a significant exclusive
decay
channels to higher K meson resonances. Besides $B \rightarrow K^i
J/\psi
(\psi^{\prime})$ processes, $B \rightarrow K^i \chi_{1c}$ channels
must have a
sizable fraction indicated by the inclusive measurement $Br(B
\rightarrow
X_s\chi_{1c}) =0.54\%$, which is even larger than $Br(B\rightarrow
X_s\psi^{\prime})
=0.30 \%$.
 The fact that the experimental data on
these exclusive decays will be available soon has been the motivation
of this
work.  In this paper, we use the heavy quark symmetries to calculate
the
hadronic matrix elements relevant to B decays to various K meson
resonances.
Also,  the model of Isgur, Wise, Scora and Grinstein \cite{IWSG} is
used for the
evaluation of the universal functions.

In section 2, we write down the effective Hamiltonian relevant to CKM
favored
nonleptonic two body decays of B meson followed by the calculation of
the
decay rates in terms of the Isgur-Wise functions.  Section 3 is
devoted to
evaluation of these universal functions using IWSG model leading to
our
estimates for various branching ratios.  We end this paper with
concluding
remarks in section 4.

\section{Effective Hamiltonian and Decay Rates}

We start with the  effective Hamiltonian relevant to $B \rightarrow
K^i
\psi (\psi^\prime , \chi_{1c})$ processes which, assuming
factorization, can be written as
\cite{AL,DTP}:
\begin{equation}
 H _{eff} = C  f_{\psi , \psi^\prime , \chi_{1c}} \bar s \gamma
_{\mu}
 (1- \gamma _5 )b \epsilon ^{\mu} _{\psi , \psi^\prime , \chi_{1c}}.
\end{equation}
 where $$ f_{\psi , \psi^\prime, \chi_{1c}} \epsilon ^{\mu} _{\psi
   , \psi^\prime , \chi_{1c}} = <0| \bar c \gamma ^{\mu}
 c| \psi , \psi^\prime , \chi_{1c}  > ,$$ \\
\begin{equation}
 C= {{G _F} \over {\sqrt 2}} (c _1 + c _2 /3 ) V _{cs} ^* V _{cb}.
\end{equation}
$c_1$ and $c_2$ are QCD improved Wilson coefficients:
\begin{equation}
{\displaystyle
\begin{array}{l}
\displaystyle{c_1 = \frac {1}{2} \left ( {\left [ \frac {\alpha _s
( \mu )}{\alpha _s ( m _W)} \right ] }^{-6/23} - {\left [ \frac
{\alpha _s ( \mu )}{\alpha _s ( m _W)} \right ] }^{12/23} \right )},
\\
\displaystyle{c_2 = \frac {1}{2} \left ( {\left [ \frac {\alpha _s
( \mu )}{\alpha _s ( m _W)} \right ] }^{-6/23} + {\left [ \frac
{\alpha _s ( \mu )}{\alpha _s ( m _W)} \right ] }^{12/23} \right )}.
\end{array}
}
\end{equation}

In the limit where {\bf b} and {\bf s} quarks are considered heavy,
matrix
elements of (1) are calculated by taking a trace:

\begin{equation}
\begin{array}{l}
<K^i (v^{\prime} )| \bar s \gamma _{\mu} (1- \gamma _5 )b |B(v)> = \\
Tr \left [\bar \Re ^i ( v^{\prime}) \gamma _ \mu (1- \gamma _5 )
\Re (v) M (v, v^{\prime}) \right ].
\end{array}
\end{equation}

$\Re ^i ( v^{\prime})$ and $\Re  ( v)$ are the matrix representations
of $K^i$ and $B$ respectively, $v$ and $v^{\prime}$ are velocities of
initial and final state mesons \cite{FALK}.  $M$ which represents the
light degrees of freedom is related to Isgur-Wise functions.  In our
previous
paper, heavy quark symmetries have been applied to the  exclusive $B
\rightarrow
(K, K^*)+( \psi , \psi ^{\prime})$ decays, in which the universal
function
has been determined by the best fit to semileptonic decay data \cite
{AL}.
In order to apply the same method when B decays to higher excited
states
of K meson, we classify these resonances into spin doublets \cite{
ATL, AOM}.
  Consequently, the decay to each spin doublet, to the leading order
in
heavy quark expansion, is expressed in terms of a single Isgur-Wise
function.
The decay rates, obtained in this way, can be written as follows:

\begin{equation}
\begin{array}{l}
\Gamma (B \rightarrow K^i \psi , \psi^\prime , \chi_{1c} ) =
\displaystyle
{\frac {C^2 f^2 _{\psi , \psi^\prime , \chi_{1c}}}{16
\pi } g(m _B , m _{K^i} , m _{\psi , \psi^\prime , \chi_{1c}} )   m
_{K^i}
F_i (x _i) {\vert \xi _I
(x _i) \vert }^2 },\\
\displaystyle{g(m _B , m _{K^i} , m _{\psi , \psi^\prime , \chi_{1c}}
) =
{\left [ {\left (1- \frac
{m^2 _{\psi , \psi^\prime , \chi_{1c}} }{m^2 _B} -\frac {m^2
_{K^i}}{m^2 _B}
\right )}^2 - \frac
{4 m^2 _{K^i} m^2 _{\psi , \psi^\prime , \chi_{1c}} }{m^4 _B} \right
]}^
{1/2}} ,\\
\displaystyle{x _i = v. v ^{\prime} = \frac {m^2 _B + m^2 _{K^i} -
m^2 _{\psi , \psi^\prime , \chi_{1c}} }{2 m _B m _{Ki} }} .
\end{array}
\end{equation}
The Isgur-Wise functions for each spin doublet $\xi _I (x) , I=
C,E,F,G$
are labeled following reference \cite{AOM}.  In table 1, the
functions $F_i(x)
$ for various excited states of K meson are tabulated.

On the other hand, using (1) we can obtain the inclusive decay rate
$\Gamma
(B \rightarrow X_s \psi , \psi^\prime , \chi_{1c} )$ which is taken
to be
 equal to $\Gamma (b \rightarrow s \psi , \psi^\prime , \chi_{1c} ) $:
\begin{equation}
\begin{array}{l}
\Gamma (b \rightarrow s \psi , \psi^\prime , \chi_{1c} )  =
\displaystyle
{\frac {C^2 f^2 _{\psi , \psi^\prime , \chi_{1c}} }
{8 \pi m_b m^2 _{\psi , \psi^\prime , \chi_{1c}}} g(m _b , m _s ,
m _{\psi , \psi^\prime , \chi_{1c}})} \times \\
\displaystyle { [m^2 _b (m^2 _b +
m^2 _{\psi , \psi^\prime , \chi_{1c}}) - m^2 _s (2 m^2 _b - m^2 _{\psi
 , \psi^\prime , \chi_{1c}} ) + m^4 _s -2 m^4 _{\psi , \psi^\prime ,
\chi_{1c}
 }]},
\end{array}
\end{equation}
where $\psi , \psi^\prime {\rm and} \chi_{1c}$ in these decays are
directly produced.  Consequently, the experimental knowledge of these
inclusive
 decays yields the branching ratio for various exclusive channels if
the
Isgur-Wise functions are known.

\section{IWSG Model for Calculating Isgur-Wise Functions}
To obtain a numerical estimate of the branching ratios, we have to
insert
in (5) the Isgur-Wise functions evaluated at $x_i$.  These functions
which represent the nonperturbative QCD effects are a measure of
light cloud
(spectator quark) rearrangement around the heavy quark during the
weak
transition.  There are various models for determining these universal
functions
from which we use the wavefunction model of Isgur, Wise, Scora and
Grinstein
\cite{IWSG}.  In this model, the functions $\xi_I$ can be
obtained from the overlap integrals
\begin{equation}
\xi (v.v^\prime )= \sqrt {2L+1} i^L \int r^2 dr \Phi^* _F (r) \Phi_I
(r)
j_L \left [ \Lambda r \sqrt {{(v.v^\prime )}^2 -1} \right ],
\end{equation}
where I and F refer to the radial wavefunction of the initial and
final state
 mesons respectively, L is the orbital angular momentum of the final
state meson
  and $j_L$ is the spherical Bessel function of order L.  The inertia
parameter
  is taken to be \cite{ALTO}
  $$\Lambda = \frac{m_{K^i}m_q}{m_s+m_q},$$
  where $m_q$ is the light quark mass.

  To calculate (7), we follow IWSG model in using harmonic oscillator
wavefunctions
   with oscillator strength $\beta$ for the radial wavefunctions
$\Phi_I$ and
    $\Phi_F$.  Note that we assume $\beta$ to be the same for initial
and final
     state mesons which is necessary if the normalization condition
$\xi_C (1
     )=1$ and $\xi_{E,F,G} (1)=0$ are to be satisfied.   For example,
using the ground state
radial wavefunction
$$\Phi (r) = \frac{\beta^{3/2}}{\pi^{3/4}} exp \left [-1/2 \beta^2 r^2
\right ],$$
one obtains
\begin{equation}
\xi_C (v.v^\prime ) = exp \left [- \frac{9}{256 \beta^2} m_{K^i} ^2
(1- {(v. v^\prime )}^2) \right ].
\end{equation}
We fix a value for $\beta$ by the best fit of (8) to
experimentally measured $B \rightarrow (K, K^* )+( \psi , \psi^\prime
)$ decays \cite{HFC}. This leads to $\beta = 0.295 GeV$ which is
smaller than
those values fitted to the semileptonic B and D decays i.e., $\beta_K
=0.34$ and $\beta _B =0.41 GeV $ \cite{IWSG}. In our previous paper
\cite{AL},
we used the same best fit of the Isgur-Wise fuction to estimate the
exclusive
decay rate for
 $B \rightarrow  K^* + \gamma$. The result is in reasonable
agreement with the recent measurment reported by CLEO II
\cite{CLEO}.  Inserting the
resulting values for the Isgur-Wise functions in (5) and using the
following experimental input \cite{HFC}
$$BR(B \rightarrow X_s \psi )=(1.09\pm 0.04\pm 0.01)\%,$$
$$BR(B \rightarrow X_s \psi^\prime )=(0.30\pm 0.05\pm0.03)\% ,$$
and \cite{1P}
$$BR(B \rightarrow X_s \chi_{1c})=(0.54\pm 0.21)\%, $$
lead to our estimate of the branching ratios reflected in tables 2, 3
and 4.
The mass parameters that appear in our calculations are taken as (in
$GeV$)
$$\begin{array}{llll}
m_q =0.33  &  m_s =0.55 \\
m_b=4.9   &  m_B =5.28 \\
m_{\psi} = 3.10  &  m_{\psi^\prime}=3.69 \\
m_{\chi_{1c}}=3.51 &
\end{array}
$$

\section{Conclusion}
We observe from Table 2 that a substantial fraction of $B \rightarrow
X_s \psi $ decays result in higher K resonances other than $K$ and
$K^*$.  For example, B decay to $K_2^* (1430)$ is comparable to its
decay to $K^*(892)$.  This is not the case, more for $B \rightarrow
X_s \psi^\prime$ and to a lesser extent for $B \rightarrow X_s
\chi_{1c}$, mainly due to phase space suppression. A large branching
ratio
for the $K_2^*(1430)$ channel in the radiative rare $B$-decays is
also pointed
out by authors of refs. \cite{ALTO,DLT,AOM}. For $B \rightarrow
K(K^*)+(c\bar c)$ prosesses, the $\chi_{1c}$-channel gives a
considerable
contribution. For example, $Br(B \rightarrow K^* +\chi_{1c})$ is even
larger than
$Br(B\rightarrow K^* +\psi^{\prime})$. Thus, for the sake of
increasing
statistics, $B \rightarrow K^i +\chi_{1c}$ channels are also favored.
The test of
these theoretical predictions should be within experimental reach
very soon.  Finally, we would like to emphasize that even though {\bf
s} quark is not particularly heavy, our results can be taken as an
order of magnitude estimate.

\noindent
{\bf Acknowledgement}

\noindent
The authors thank A. Ali and R. R. Mendel for useful discussions.

\newpage

\hskip -1cm
{\small
\begin{tabular}{c|c|c}
\hline
\multicolumn{1}{|c|}{$K^i$ Name} &
\multicolumn{1}{c|}{$ J^P $} &
\multicolumn{1}{c|}{$F_i  (x)$} \cr
\hline
\multicolumn{1}{|c|}{$K (498)$} &
\multicolumn{1}{c|}{$0^-$} &
\multicolumn{1}{c|}{$(x+1)[(x+1){(m_B -m_K)}^2/m_{\psi} ^2-2]$} \cr
\hline
\multicolumn{1}{|c|}{$K^* (892)$} &
\multicolumn{1}{c|}{$1^-$} &
\multicolumn{1}{c|}{$(x+1)[(x-1){(m_B +m_{K^*})}^2/m_{\psi} ^2+4x+2]$}
\cr
\hline
\multicolumn{1}{|c|}{$K^* (1430)$} &
\multicolumn{1}{c|}{$0^+$} &
\multicolumn{1}{c|}{$(x-1)[(x-1){(m_B +m_{K^*})}^2/m_{\psi} ^2+2]$}
\cr
\hline
\multicolumn{1}{|c|}{$K _1 (1270)$} &
\multicolumn{1}{c|}{$1^+$} &
\multicolumn{1}{c|}{$(x-1)[(x+1){(m_B -m_{K_1})}^2/m_{\psi} ^2+4x-2]$}
\cr
\hline
\multicolumn{1}{|c|}{$K_1 (1400)$} &
\multicolumn{1}{c|}{$1^+$} &
\multicolumn{1}{c|}{$2/3(x-1){(x+1)}^2[(x+1){(m_B -
m_{K_1})}^2/m_{\psi}
^2+x-2]$} \cr
\hline
\multicolumn{1}{|c|}{$K^* _2 (1430)$} &
\multicolumn{1}{c|}{$2^+$} &
\multicolumn{1}{c|}{$2/3(x-1){(x+1)}^2[(x-1){(m_B +m_{K^* _2})}^2/
m_{\psi} ^2+3x+2]$} \cr
\hline
\multicolumn{1}{|c|}{$K^* (1680)$} &
\multicolumn{1}{c|}{$1^-$} &
\multicolumn{1}{c|}{$2/3(x+1){(x-1)}^2[{(x-1)}^2{(m_B
+m_{K^*})}^2/m_{\psi}
^2+x+2]$} \cr
\hline
\multicolumn{1}{|c|}{$K_2 (1580)$} &
\multicolumn{1}{c|}{$2^-$} &
\multicolumn{1}{c|}{$2/3(x+1){(x-1)}^2[(x+1){(m_B -
m_{K_2})}^2/m_{\psi}
^2+3x-2]$} \cr
\hline
\multicolumn{1}{|c|}{$K(1460)$} &
\multicolumn{1}{c|}{$0^-$} &
\multicolumn{1}{c|}{$(x+1)[(x+1){(m_B -m_K)}^2/m_{\psi} ^2-2]$} \cr
\hline
\multicolumn{1}{|c|}{$K^* (1410)$} &
\multicolumn{1}{c|}{$1^-$} &
\multicolumn{1}{c|}{$(x+1)[(x-1){(m_B +m_{K^*})}^2/m_{\psi} ^2+4x+2]$}
\cr
\hline
\end{tabular}
\hskip 1cm
\center{Table 1. $F _i (x)$ for various K-meson
excited states. }}

\vskip 2cm
\hskip -1cm
{\small
\begin{tabular}{c|c|c|c|c|c}
\hline
\multicolumn{1}{|c|}{$K^i$ Name} &
\multicolumn{1}{c|}{$ J^P $} &
\multicolumn{1}{c|}{Mass (MeV)} &
\multicolumn{1}{c|}{$ x _{\circ}$} &
\multicolumn{1}{c|}{$\xi _I (x_{\circ})$} &
\multicolumn{1}{c|}{$\begin{array}{l} BR(B \rightarrow K^i \psi
)\end{array}$}
\cr
\hline
\multicolumn{1}{|c|}{$K $} &
\multicolumn{1}{c|}{$0^-$} &
\multicolumn{1}{c|}{$497.67 \pm 0.03$} &
\multicolumn{1}{c|}{3.52} &
\multicolumn{1}{c|}{0.319} &
\multicolumn{1}{c|}{0.094} \cr
\hline
\multicolumn{1}{|c|}{$K^* (892)$} &
\multicolumn{1}{c|}{$1^-$} &
\multicolumn{1}{c|}{$896.1 \pm 0.3$} &
\multicolumn{1}{c|}{2.02} &
\multicolumn{1}{c|}{0.368} &
\multicolumn{1}{c|}{0.227} \cr
\hline
\multicolumn{1}{|c|}{$K^* (1430)$} &
\multicolumn{1}{c|}{$0^+$} &
\multicolumn{1}{c|}{$1429 \pm 7$} &
\multicolumn{1}{c|}{1.35} &
\multicolumn{1}{c|}{0.631} &
\multicolumn{1}{c|}{0.026} \cr
\hline
\multicolumn{1}{|c|}{$K _1 (1270)$} &
\multicolumn{1}{c|}{$1^+$} &
\multicolumn{1}{c|}{$1270 \pm 10$} &
\multicolumn{1}{c|}{1.48} &
\multicolumn{1}{c|}{0.626} &
\multicolumn{1}{c|}{0.075} \cr
\hline
\multicolumn{1}{|c|}{$K_1 (1400)$} &
\multicolumn{1}{c|}{$1^+$} &
\multicolumn{1}{c|}{$1402 \pm 7$} &
\multicolumn{1}{c|}{1.37} &
\multicolumn{1}{c|}{0.630} &
\multicolumn{1}{c|}{0.087} \cr
\hline
\multicolumn{1}{|c|}{$K^* _2 (1430)$} &
\multicolumn{1}{c|}{$2^+$} &
\multicolumn{1}{c|}{$1425.4 \pm 1.3$} &
\multicolumn{1}{c|}{1.35} &
\multicolumn{1}{c|}{0.631} &
\multicolumn{1}{c|}{0.202} \cr
\hline
\multicolumn{1}{|c|}{$K^* (1680)$} &
\multicolumn{1}{c|}{$1^-$} &
\multicolumn{1}{c|}{$1714 \pm 20$} &
\multicolumn{1}{c|}{1.17} &
\multicolumn{1}{c|}{0.326} &
\multicolumn{1}{c|}{0.001} \cr
\hline
\multicolumn{1}{|c|}{$K_2 (1580)$} &
\multicolumn{1}{c|}{$2^-$} &
\multicolumn{1}{c|}{$\approx 1580$} &
\multicolumn{1}{c|}{1.24} &
\multicolumn{1}{c|}{0.364} &
\multicolumn{1}{c|}{0.003} \cr
\hline
\multicolumn{1}{|c|}{$K(1460)$} &
\multicolumn{1}{c|}{$0^-$} &
\multicolumn{1}{c|}{$\approx 1460$} &
\multicolumn{1}{c|}{1.32} &
\multicolumn{1}{c|}{-0.275} &
\multicolumn{1}{c|}{0.014} \cr
\hline
\multicolumn{1}{|c|}{$K^* (1410)$} &
\multicolumn{1}{c|}{$1^-$} &
\multicolumn{1}{c|}{$1412 \pm 12$} &
\multicolumn{1}{c|}{1.36} &
\multicolumn{1}{c|}{-0.282} &
\multicolumn{1}{c|}{0.088} \cr
\hline
\multicolumn{1}{|c|}{} &
\multicolumn{1}{c|}{} &
\multicolumn{1}{c|}{} &
\multicolumn{1}{c|}{} &
\multicolumn{1}{c|}{} &
\multicolumn{1}{c|}{TOTAL: 0.817} \cr
\hline
\end{tabular}
\hskip 1cm
\center{Table 2: The branching ratios for B decay to $\psi$ and
various K mesons.}}

\vskip 2cm
\hskip -1cm
{\small
\begin{tabular}{c|c|c|c|c|c|c}
\hline
\multicolumn{1}{|c|}{$K^i$ Name} &
\multicolumn{1}{c|}{$ J^P $} &
\multicolumn{1}{c|}{Mass (MeV)} &
\multicolumn{1}{c|}{$ x _{\circ}$} &
\multicolumn{1}{c|}{$\xi _I (x_{\circ})$} &
\multicolumn{1}{c|}{$\begin{array}{l} BR(B \rightarrow K^i
\psi^\prime
)\end{array}$} \cr
\hline
\multicolumn{1}{|c|}{$K $} &
\multicolumn{1}{c|}{$0^-$} &
\multicolumn{1}{c|}{$497.67 \pm 0.03$} &
\multicolumn{1}{c|}{2.76} &
\multicolumn{1}{c|}{0.515} &
\multicolumn{1}{c|}{0.052} \cr
\hline
\multicolumn{1}{|c|}{$K^* (892)$} &
\multicolumn{1}{c|}{$1^-$} &
\multicolumn{1}{c|}{$896.1 \pm 0.3$} &
\multicolumn{1}{c|}{1.59} &
\multicolumn{1}{c|}{0.609} &
\multicolumn{1}{c|}{0.180} \cr
\hline
\multicolumn{1}{|c|}{$K^* (1430)$} &
\multicolumn{1}{c|}{$0^+$} &
\multicolumn{1}{c|}{$1429 \pm 7$} &
\multicolumn{1}{c|}{1.08} &
\multicolumn{1}{c|}{0.433} &
\multicolumn{1}{c|}{0.001} \cr
\hline
\multicolumn{1}{|c|}{$K _1 (1270)$} &
\multicolumn{1}{c|}{$1^+$} &
\multicolumn{1}{c|}{$1270 \pm 10$} &
\multicolumn{1}{c|}{1.18} &
\multicolumn{1}{c|}{0.538} &
\multicolumn{1}{c|}{0.005} \cr
\hline
\multicolumn{1}{|c|}{$K_1 (1400)$} &
\multicolumn{1}{c|}{$1^+$} &
\multicolumn{1}{c|}{$1402 \pm 7$} &
\multicolumn{1}{c|}{1.10} &
\multicolumn{1}{c|}{0.466} &
\multicolumn{1}{c|}{0.002} \cr
\hline
\multicolumn{1}{|c|}{$K^* _2 (1430)$} &
\multicolumn{1}{c|}{$2^+$} &
\multicolumn{1}{c|}{$1425.4 \pm 1.3$} &
\multicolumn{1}{c|}{1.08} &
\multicolumn{1}{c|}{0.432} &
\multicolumn{1}{c|}{0.004} \cr
\hline
\multicolumn{1}{|c|}{$K^* (1680)$} &
\multicolumn{1}{c|}{$1^-$} &
\multicolumn{1}{c|}{$1714 \pm 20$} &
\multicolumn{1}{c|}{0.950} &
\multicolumn{1}{c|}{-} &
\multicolumn{1}{c|}{-} \cr
\hline
\multicolumn{1}{|c|}{$K_2 (1580)$} &
\multicolumn{1}{c|}{$2^-$} &
\multicolumn{1}{c|}{$\approx 1580$} &
\multicolumn{1}{c|}{1.00} &
\multicolumn{1}{c|}{-} &
\multicolumn{1}{c|}{-} \cr
\hline
\multicolumn{1}{|c|}{$K(1460)$} &
\multicolumn{1}{c|}{$0^-$} &
\multicolumn{1}{c|}{$\approx 1460$} &
\multicolumn{1}{c|}{1.06} &
\multicolumn{1}{c|}{-0.078} &
\multicolumn{1}{c|}{$<$ 0.001} \cr
\hline
\multicolumn{1}{|c|}{$K^* (1410)$} &
\multicolumn{1}{c|}{$1^-$} &
\multicolumn{1}{c|}{$1412 \pm 12$} &
\multicolumn{1}{c|}{1.09} &
\multicolumn{1}{c|}{-0.106} &
\multicolumn{1}{c|}{0.003} \cr
\hline
\multicolumn{1}{|c|}{} &
\multicolumn{1}{c|}{} &
\multicolumn{1}{c|}{} &
\multicolumn{1}{c|}{} &
\multicolumn{1}{c|}{} &
\multicolumn{1}{c|}{TOTAL: 0.247} \cr
\hline
\end{tabular}
\hskip 1cm
\center{Table 3: The branching ratios for B decay to $\psi^\prime$
and various
K mesons.}}

\vskip 2cm
\hskip -1cm
{\small
\begin{tabular}{c|c|c|c|c|c|c}
\hline
\multicolumn{1}{|c|}{$K^i$ Name} &
\multicolumn{1}{c|}{$ J^P $} &
\multicolumn{1}{c|}{Mass (MeV)} &
\multicolumn{1}{c|}{$ x _{\circ}$} &
\multicolumn{1}{c|}{$\xi _I (x_{\circ})$} &
\multicolumn{1}{c|}{$\begin{array}{l} BR(B \rightarrow K^i \chi_{1c}
)\end{array}$} \cr
\hline
\multicolumn{1}{|c|}{$K $} &
\multicolumn{1}{c|}{$0^-$} &
\multicolumn{1}{c|}{$497.67 \pm 0.03$} &
\multicolumn{1}{c|}{3.01} &
\multicolumn{1}{c|}{0.446} &
\multicolumn{1}{c|}{0.076} \cr
\hline
\multicolumn{1}{|c|}{$K^* (892)$} &
\multicolumn{1}{c|}{$1^-$} &
\multicolumn{1}{c|}{$896.1 \pm 0.3$} &
\multicolumn{1}{c|}{1.73} &
\multicolumn{1}{c|}{0.524} &
\multicolumn{1}{c|}{0.235} \cr
\hline
\multicolumn{1}{|c|}{$K^* (1430)$} &
\multicolumn{1}{c|}{$0^+$} &
\multicolumn{1}{c|}{$1429 \pm 7$} &
\multicolumn{1}{c|}{1.17} &
\multicolumn{1}{c|}{0.565} &
\multicolumn{1}{c|}{0.004} \cr
\hline
\multicolumn{1}{|c|}{$K _1 (1270)$} &
\multicolumn{1}{c|}{$1^+$} &
\multicolumn{1}{c|}{$1270 \pm 10$} &
\multicolumn{1}{c|}{1.28} &
\multicolumn{1}{c|}{0.605} &
\multicolumn{1}{c|}{0.020} \cr
\hline
\multicolumn{1}{|c|}{$K_1 (1400)$} &
\multicolumn{1}{c|}{$1^+$} &
\multicolumn{1}{c|}{$1402 \pm 7$} &
\multicolumn{1}{c|}{1.18} &
\multicolumn{1}{c|}{0.568} &
\multicolumn{1}{c|}{0.010} \cr
\hline
\multicolumn{1}{|c|}{$K^* _2 (1430)$} &
\multicolumn{1}{c|}{$2^+$} &
\multicolumn{1}{c|}{$1425.4 \pm 1.3$} &
\multicolumn{1}{c|}{1.17} &
\multicolumn{1}{c|}{0.564} &
\multicolumn{1}{c|}{0.032} \cr
\hline
\multicolumn{1}{|c|}{$K^* (1680)$} &
\multicolumn{1}{c|}{$1^-$} &
\multicolumn{1}{c|}{$1714 \pm 20$} &
\multicolumn{1}{c|}{1.02} &
\multicolumn{1}{c|}{0.053} &
\multicolumn{1}{c|}{$<$ 0.001} \cr
\hline
\multicolumn{1}{|c|}{$K_2 (1580)$} &
\multicolumn{1}{c|}{$2^-$} &
\multicolumn{1}{c|}{$\approx 1580$} &
\multicolumn{1}{c|}{1.08} &
\multicolumn{1}{c|}{0.164} &
\multicolumn{1}{c|}{$<$ 0.001} \cr
\hline
\multicolumn{1}{|c|}{$K(1460)$} &
\multicolumn{1}{c|}{$0^-$} &
\multicolumn{1}{c|}{$\approx 1460$} &
\multicolumn{1}{c|}{1.15} &
\multicolumn{1}{c|}{-0.172} &
\multicolumn{1}{c|}{0.001} \cr
\hline
\multicolumn{1}{|c|}{$K^* (1410)$} &
\multicolumn{1}{c|}{$1^-$} &
\multicolumn{1}{c|}{$1412 \pm 12$} &
\multicolumn{1}{c|}{1.18} &
\multicolumn{1}{c|}{-0.188} &
\multicolumn{1}{c|}{0.017} \cr
\hline
\multicolumn{1}{|c|}{} &
\multicolumn{1}{c|}{} &
\multicolumn{1}{c|}{} &
\multicolumn{1}{c|}{} &
\multicolumn{1}{c|}{} &
\multicolumn{1}{c|}{TOTAL: 0.395} \cr
\hline
\end{tabular}
\hskip 1cm
\center{Table 4: The branching ratios for B decay to $\chi_{1c}$ and
various K mesons.}}

\end{document}